\documentclass[10pt,journal,compsoc]{IEEEtran}

\IEEEoverridecommandlockouts

\usepackage{cite}
\usepackage{booktabs} 
\usepackage{graphicx}
\usepackage{array}
\usepackage{url}
\usepackage{fancybox}
\usepackage{multirow}
\usepackage{amsmath,amssymb,amsfonts}
\usepackage{color}
\usepackage[table,xcdraw]{xcolor}
\usepackage{fancyhdr}
\usepackage{subfig}
\usepackage{diagbox}
\usepackage{bbding}
\usepackage{placeins}
\usepackage{balance}
\usepackage{hyperref} 
\usepackage{enumitem}
\usepackage[linesnumbered, ruled,vlined]{algorithm2e}
\def\BibTeX{{\rm B\kern-.05em{\sc i\kern-.025em b}\kern-.08em
		T\kern-.1667em\lower.7ex\hbox{E}\kern-.125emX}}

\usepackage{listings, xcolor}
\usepackage{url}
\usepackage{tabularray}
\usepackage{tcolorbox}
\usepackage{amsmath}
\usepackage{listings}
\usepackage{tikz}
\usepackage{multirow}
\usepackage{tabularx}
\usepackage{arydshln}
\definecolor{NGreen}{RGB}{0,176,80}
\definecolor{verylightgray}{rgb}{.97,.97,.97}
\lstdefinelanguage{Solidity}{
  keywords=[1]{anonymous, assembly, assert, balance, break, call, callcode, case, catch, class, constant, continue, constructor, contract, debugger, default, delegatecall, delete, do, else, emit, event, experimental, export, external, false, finally, for, function, gas, if, implements, import, in, indexed, instanceof, interface, internal, is, length, library, log0, log1, log2, log3, log4, memory, modifier, new, payable, pragma, private, protected, public, pure, push, require, return, returns, revert, selfdestruct, send, solidity, storage, struct, suicide, super, switch, then, this, throw, transfer, true, try, typeof, using, value, view, while, with, addmod, ecrecover, keccak256, mulmod, ripemd160, sha256, sha3}, 
  keywordstyle=[1]\color{blue}\bfseries,
  keywords=[2]{address, bool, byte, bytes, bytes1, bytes2, bytes3, bytes4, bytes5, bytes6, bytes7, bytes8, bytes9, bytes10, bytes11, bytes12, bytes13, bytes14, bytes15, bytes16, bytes17, bytes18, bytes19, bytes20, bytes21, bytes22, bytes23, bytes24, bytes25, bytes26, bytes27, bytes28, bytes29, bytes30, bytes31, bytes32, enum, int, int8, int16, int24, int32, int40, int48, int56, int64, int72, int80, int88, int96, int104, int112, int120, int128, int136, int144, int152, int160, int168, int176, int184, int192, int200, int208, int216, int224, int232, int240, int248, int256, mapping, string, uint, uint8, uint16, uint24, uint32, uint40, uint48, uint56, uint64, uint72, uint80, uint88, uint96, uint104, uint112, uint120, uint128, uint136, uint144, uint152, uint160, uint168, uint176, uint184, uint192, uint200, uint208, uint216, uint224, uint232, uint240, uint248, uint256, var, void, ether, finney, szabo, wei, days, hours, minutes, seconds, weeks, years},  
  keywordstyle=[2]\color{teal}\bfseries,
  keywords=[3]{block, blockhash, coinbase, difficulty, gaslimit, number, timestamp, msg, data, gas, sender, sig, value, now, tx, gasprice, origin},  
  keywordstyle=[3]\color{violet}\bfseries,
  identifierstyle=\color{black},
  sensitive=false,
  comment=[l]{//},
  morecomment=[s]{/*}{*/},
  commentstyle=\color{gray}\ttfamily,
  stringstyle=\color{red}\ttfamily,
  morestring=[b]',
  morestring=[b]"
}
\lstdefinestyle{datalogstyle}{
  language=Datalog,
  basicstyle=\ttfamily\small,
  keywordstyle=\bfseries,
  commentstyle=\itshape,
  morekeywords={:-},
  deletekeywords={not},
  literate=
    {:-}{{\textcolor{blue}{:-}}}2
    {,}{{\textcolor{red}{,}}}1
    {.}{{\textcolor{red}{.}}}1
}

\begin{document}
	
	\title{CRPWarner: Warning the Risk of Contract-related Rug Pull in DeFi Smart Contracts}
				
   
	\author{Zewei Lin, Jiachi Chen, Zibin Zheng,~\IEEEmembership{Fellow,~IEEE,}, Jiajing Wu, Weizhe Zhang,  Yongjuan Wang
		\IEEEcompsocitemizethanks{\IEEEcompsocthanksitem Zewei Lin, Jiachi Chen, Zibin Zheng, Jiajing Wu are with School of Software Engineering, Sun Yat-sen University, China. \protect\\
                E-mail: linzw3@mail2.sysu.edu.cn
			E-mail: \{chenjch86, zhzibin, wujiajing\}@mail.sysu.edu.cn

            \IEEEcompsocthanksitem Weizhe Wang is with the School of Computer Science and Technology, Harbin Institute of Technology, China.\protect\\
			E-mail: wzzhang@hit.edu.cn 

            \IEEEcompsocthanksitem Yongjuan Wang is with the Henan Key Laboratory of Network Cryptography Technology, China.\protect\\
			E-mail: pinkywyj@163.com
			
			\IEEEcompsocthanksitem Zibin Zheng is the corresponding author.}
		\thanks{Manuscript received     ; revised   }}

	\markboth{IEEE Transactions on Software Engineering, ~Vol.~  , No.~  , }%
	{Shell \MakeLowercase{\textit{et al.}}: Bare Demo of IEEEtran.cls for Computer Society Journals}

	\IEEEtitleabstractindextext{%
            \begin{abstract}
In recent years, Decentralized Finance (DeFi) grows rapidly due to the development of blockchain technology and smart contracts. As of March 2023, the estimated global cryptocurrency market cap has reached approximately \$949 billion. However, security incidents continue to plague the DeFi ecosystem, and one of the most notorious examples is the ``Rug Pull'' scam. This type of cryptocurrency scam occurs when the developer of a particular token project intentionally abandons the project and disappears with investors' funds. Despite it only emerging in recent years, Rug Pull events have already caused significant financial losses.

In this work, we manually collected and analyzed 103 real-world rug pull events, categorizing them based on their scam methods. Two primary categories were identified: {\itshape Contract-related} Rug Pull (through malicious functions in smart contracts) and {\itshape Transaction-related} Rug Pull (through cryptocurrency trading without utilizing malicious functions). Based on the analysis of rug pull events, we propose CRPWarner (short for \textbf{C}ontract-related \textbf{R}ug \textbf{P}ull Risk \textbf{Warner}) to identify malicious functions in smart contracts and issue warnings regarding potential rug pulls. We evaluated CRPWarner on 69 open-source smart contracts related to rug pull events and achieved a 91.8\% precision, 85.9\% recall and 88.7\% F1-score. Additionally, when evaluating CRPWarner on 13,484 real token contracts on Ethereum, it successfully detected 4168 smart contracts with malicious functions, including zero-day examples. The precision of large-scale experiment reach 84.9\%.
\end{abstract}

	\begin{IEEEkeywords}
			Smart Contracts, Decentralized Finance, Rug Pull, Datalog Analysis
	\end{IEEEkeywords}
        }

	
	\maketitle
	\IEEEdisplaynontitleabstractindextext


	\section{Introduction}
\label{Introduction}

With the rapid development of blockchain and smart contract technologies, an increasing number of applications are being built upon them. 
One of the most popular applications based on blockchain technology is Web3~\cite{buldas2022towards, buldas2022ultra}, representing a decentralized internet ecosystem that is both owned and operated by its users~\cite{web3chance}. Web3 encompasses various elements such as cryptocurrencies, Non-Fungible Tokens (NFTs), and Decentralized Finance (DeFi)~\cite{web3what}. DeFi, in particular, constitutes a financial system functioning independently of centralized third-party institutions~\cite{werner2021sok}. 
The DeFi ecosystem offers a wide range of financial services, such as cryptocurrency trading and uncollateralized loans. As of March 12, 2023, the global crypto market cap is estimated to be approximately \$949 billion \cite{CoinMarketCap}. Unfortunately, the DeFi ecosystem has experienced numerous security incidents, including front-running \cite{daian2020flash}, flash loan attacks \cite{qin2021attacking}, and rug pulls \cite{mazorra2022not}.

Rug pull is a type of cryptocurrency scam that occurs when the developer of a particular token project intentionally abandons the project and vanishes with investors' funds. This makes it difficult for investors to recover their invested funds by selling the cryptocurrencies they hold, as these cryptocurrencies become worthless. Though rug pulls have only emerged in recent years, they have already caused significant financial losses. In fact, rug pulls account for 37\% of all cryptocurrency scam losses in 2021, compared to mere 1\% in 2020, making it one of the largest scam types in the DeFi ecosystem. In 2021, rug pulls resulted in over \$2.8 billion in cryptocurrency losses \cite{chainalysisRugPull}. 

Existing work~\cite{mazorra2022not, xia2021trade} on analyzing Rug Pull events in DeFi projects is limited to post-event detection, lacking the capability to provide pre-warning before such events occur. Current detection methods involve labeling datasets through factors like token price changes and the similarity between DeFi project names and well-known project names. Subsequently, machine learning-based approaches are employed to analyze transaction records and identify Rug Pull events. These methodologies rely on features learned by machine learning models without conducting a thorough analysis of the causes behind the Rug Pull events. Moreover, it necessitates a substantial volume of transaction records for DeFi projects, making it impractical for early warning systems upon project launch.

In this work, we manually collected real-world rug pull events, systematically analyzed and classified them based on their scam methods. We gathered the rug pull events from several blockchain security platforms~\cite{Rugdoc,PeckShieldAlert,Slowmist} and conducted a comprehensive manual analysis. We classified them into two main categories: \textit{Contract-related} Rug Pull, which occur through malicious functions in smart contracts, and \textit{Transaction-related} Rug Pull, which occur through cryptocurrency trading without the use of malicious functions. Then, we further categorized these two types of rug pulls to provide a more detailed analysis. The \textit{Contract-related} Rug Pull can be further categorized into three types: {\itshape Hidden Mint Function}, {\itshape Limiting Sell Order} and {\itshape Leaking Token}. Meanwhile, the \textit{Transaction-related} Rug Pull can be further categorized into three types: {\itshape Dumping Cryptocurrency}, {\itshape Withdrawing Liquidity} and {\itshape Abandoning Project after Funding Completion} (see Section 3 for details). These identified patterns serve as critical indicators of risk for users, alerting them to potential threats to their assets in contracts exhibiting these patterns. For developers, this knowledge can help them avoid the inadvertent inclusion of such patterns in their contracts, as they can erode user confidence.

Based on the analysis of rug pull events, we propose CRPWarner, an automated analysis method that assesses the risk of \textit{Contract-related} Rug Pull. CRPWarner serves two primary applications. Firstly, it can provide early warnings of potential rug pulls in DeFi projects by detecting malicious functions in smart contracts that could be used to execute a rug pull. Secondly, smart contract developers can use it to detect and delete unnecessary high-risk functions, protecting the project from attacks that leverage these functions in the event of private key theft.

CRPWarner first decompile the EVM bytecode, build a control flow graph (CFG), and subsequently performs a domain-specific datalog analysis \cite{immerman2012descriptive}. The datalog analysis consists of two primary components: information flow analysis and malicious function identification. The information flow analysis involves the examination of variables and functions that have a strong association with rug pull and the corresponding semantics.

To evaluate the effectiveness of CRPWarner, we employed it to analyze two datasets. Initially, we tested CRPWarner on the 69 open-source smart contracts associated with rug pull events to assess its accuracy. The experimental results revealed that CRPWarner can detect malicious functions in smart contracts with precision, recall and F1-score of 91.8\%, 85.9\% and 88.7\%. Next, we employed CRPWarner to analyze a large-scale dataset of 13,484 real-world ERC token contracts to further evaluate its effectiveness, the precision of
the large-scale experiment reached 84.9\%. CRPWarner discovered a zero-day example: a token project with a malicious function that had not been reported.

In summary, the main contributions of this work are as follows:
\begin{itemize}
\item We collected and manually analysed the smart contracts and transaction records of 103 real-world rug pull events, and categorised them based on the specific method used for each rug pull.
\item We propose CRPWarner, an analysis tool designed to assess the risk of \textit{Contract-related} Rug Pull. We used CRPWarner to analyze open-source rug pull events and evaluated its effectiveness, achieving precision, recall and F1-score of 91.8\%, 85.9\%, and 88.7\%.
\item We release our dataset and the code of CRPWarner at the Github repository~\footnote{https://github.com/CRPWarner/RugPull}.
\end{itemize}

The remainder of this paper is organized as follows: Section 2 introduces the background related to rug pull. Section 3 presents the analysis and classification of rug pull events. Section 4 outlines the detailed design of our approach. Section 5 gives an experimental evaluation, and section 6 summarises the related work. Finally, we concludes this work in section 7.

	\section{Background}

\subsection{Smart Contract and Transactions}
Smart contracts are programs that execute automatically on the blockchain. They can be applied in a variety of scenarios, such as supply chain management \cite{de2020blockchain} and financial transactions \cite{schar2021decentralized}. Smart contracts are created and invoked though transactions, and transactions can be automatically enforced through smart contracts. Transactions on a blockchain network can be classified into two types: external and internal. External transactions involve two Externally Owned Accounts (EOAs), while internal transactions occur between a smart contract and another smart contract or EOAs.

The source code for smart contracts is written in high-level programming languages, one of which is Solidity. Bytecode is a low-level representation of code that is typically generated from source code by a compiler \cite{EVM} and stored on the blockchain for the purpose of executing smart contracts. When a Solidity-written smart contract is deployed on Ethereum \cite{Etherscan}, the bytecode is stored on the blockchain and the contract is given a unique address. To interact with a smart contract, a user can initiate the execution of bytecode by sending a transaction to the contract's address.

\subsection{Web3 and DeFi}
Web3 constitutes a decentralized, blockchain-based internet ecosystem owned and operated by its users, embodying a movement towards a more equitable and just internet~\cite{web3chance, buldas2022towards, buldas2022ultra}. Web3 offers several advantages. Firstly, it empowers users with greater control over the content they create and their digital assets. Secondly, projects within Web3 can more easily gain users' trust due to the auditable and immutable nature of blockchain and smart contracts~\cite{web3build}. An integral component of Web3 is Decentralized Finance (DeFi)~\cite{web3what}.

DeFi is a peer-to-peer financial system built on top of smart contracts \cite{werner2021sok}. DeFi, as the name suggests, does not rely on centralized intermediaries such as banks and other financial institutions. Cryptocurrency trading is one of the most popular applications of DeFi, which is known as the act of buying and selling cryptocurrencies for profit \cite{fang2022cryptocurrency}. The global crypto market cap is estimated to be around 949 billion dollars as of March, 2023 \cite{CoinMarketCap}. 

\begin{figure}[h]
    \centering
    \includegraphics[width=\linewidth]{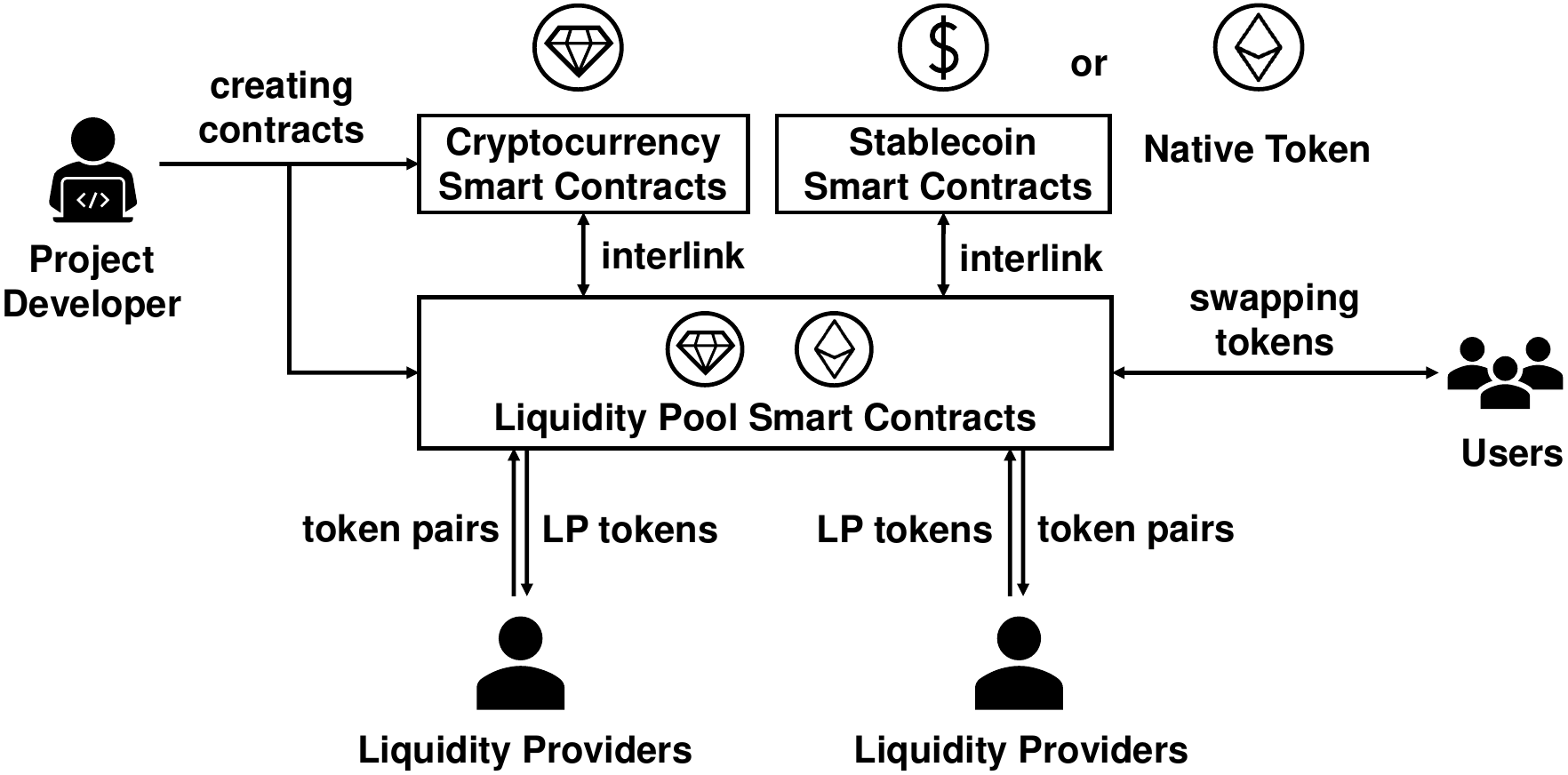}
    \caption{Progress of Cryptocurrency Trading}  
    \label{fig: crypto trading} 
\end{figure}

The most popular method of trading cryptocurrency is through automated market makers (AMMs) \cite{fang2022cryptocurrency}. These platforms enable cryptocurrency to be traded automatically by utilizing liquidity pools, rather than relying on the traditional market of buyers and sellers. Users can supply a pair of tokens to liquidity pools and receive the corresponding liquidity pool tokens (LP Tokens) in return, which is called providing liquidity. Liquidity providers can earn fees from traders who swap tokens within the liquidity pools. If necessary, they can also withdraw their funds by burning the LP tokens and taking their tokens back. This progress is illustrated in Figure \ref{fig: crypto trading}.

From the developer's perspective, creating a cryptocurrency and profiting from it requires a significant amount of technical and business expertise \cite{CreateERC20token} Using ERC20 tokens \cite{ERC20} as an example, the process typically involves two steps: the development of smart contracts, and marketing efforts aimed at generating rewards. 

\textbf{The development of smart contracts.} Firstly, developers need to create a smart contract for the token they want to issue. Once the token smart contract has been compiled and deployed on the blockchain, developers must then create liquidity pools that can be used to pair the newly created token with other stablecoins or the native token of the blockchain, allowing users to easily trade between different cryptocurrencies.

\textbf{Marketing efforts to generate rewards.} After developing smart contracts, developers need to make marketing efforts to generate rewards. For instance, they can associate a token with certain features or services, to attract investors or facilitate transactions, and then make rewards by offering transaction fees, selling tokens through an initial coin offering (ICO), or other methods.

	\section{Rug Pull Study}

A rug pull is a type of cryptocurrency scam that occurs when the developer or team behind a token project intentionally abandons the project and disappears with investors' funds \cite{RugPull}. This type of scam leaves investors with a worthless asset, making it difficult for them to recoup their investment funds by selling their holding tokens.

Rug pull has gradually evolved into one of the most financially damaging types of cryptocurrency scams in recent years \cite{chainalysisRugPull}. In this section, we manually analyzed the smart contracts and transaction records associated with rug pull events and classified them based on their attack methods.

In the existing work, rug pulls are commonly classified into two distinct types based on their reaction time: hard and soft rug pulls~\cite{hardsoft1, hardsoft2, hardsoft3}. Hard rug pulls are characterized by their abrupt and immediate nature, leading to an instantaneous and comprehensive loss of investor funds, such as a sudden withdrawal of funds from a project through the malicious functions in smart contracts~\cite{hardsoft2}. On the contrary, soft rug pulls unfold gradually over an extended period, exemplified by merely dumping their tokens while maintaining a facade of continued investment and support for the project within the community~\cite{hardsoft3}. It is noteworthy that a universally established standard to delineate the temporal threshold distinguishing hard from soft rug pulls is presently lacking, thereby impacting the precision of classification. The absence of this standard implies that the same rug pull event may be classified into different types under varying time threshold standards. To enhance classification accuracy, we propose a new method grounded in their attack mechanisms rather than reaction time.


\subsection{Analysis of Rug Pull Events}
\subsubsection{Data Collection}
To analyze and categorize rug pull events, it is important to gather reports about such events. We collected all reports on rug pull events until January 2024 from three blockchain security platforms, i.e., PeckShield~\cite{PeckShieldAlert}, SlowMist~\cite{Slowmist}, and RugDoc~\cite{Rugdoc}, utilizing the keywords ``rug pull'' and ``rug''. These rug pull events span various notable blockchain networks, encompassing Ethereum (ETH)~\cite{Etherscan}, Binance Smart Chain (BSC)~\cite{BSCscan}, Fantom blockchain (FTM)~\cite{FTM}, and others. In total, we collected 103 reports about real-world rug pull events, and we have publicized the analysis results of these events in the GitHub repository~\footnote{https://github.com/CRPWarner/RugPul}.


\subsubsection{Manaual Filtering}
In the previous subsection, we described the collection of 103 reports about real-world rug pull events. However, some of these reports lack essential information required for analyzing the attack method, such as the associated blockchain and contract address. Absent this information, obtaining transaction records or smart contract source code becomes unfeasible, hindering the analysis of the root cause behind the rug pull events. Therefore, we manually removed these reports. Ultimately, following manual filtering, we identified 93 out of 103 collected reports about rug pull events.

\subsubsection{Open Card Sorting}

\begin{table*}[h]
\centering
\setlength{\abovecaptionskip}{0.05cm}
\caption{Definition of Rug Pull Events}
\label{tab: rug pull def}
\resizebox{\linewidth}{!}{
\begin{tabular}{|cl|l|}
\hline
\multicolumn{2}{|c|}{\textbf{Rug Pull Type}}                                                             & \multicolumn{1}{c|}{\textbf{Definition}}                                 \\ \hline
\multicolumn{1}{|c|}{\multirow{3}{*}{\textit{Contract-Related}}}    & Hidden Mint Function                        & Exist a mint function to generate any number of tokens to any address.   \\ \cline{2-3} 
\multicolumn{1}{|c|}{}                                     & Limiting Sell Order                         & Exist a mechanism to restrict users from selling tokens.                 \\ \cline{2-3} 
\multicolumn{1}{|c|}{}                                     & Leaking Token                               & Exist a mechanism to leak token from other users without permission.     \\ \hline
\multicolumn{1}{|l|}{\multirow{3}{*}{\textit{Transaction-Related}}} & Dumping Cryptocurrency                      & Sell off a large number of cryptocurrency suddenly.                      \\ \cline{2-3} 
\multicolumn{1}{|l|}{}                                     & Withdrawing Liquidity                       & Withdraw liquidity and steal almost all the valuable assets in the pool. \\ \cline{2-3} 
\multicolumn{1}{|l|}{}                                     & Abandoning Project after Funding & Abandon the project and abscond without delivering on their promises.    \\ \hline
\end{tabular}}
\end{table*}

To ensure the accuracy of the results, we conducted an analysis and categorization of the filtered reports about real-world rug pull events through the open card sorting approach~\cite{spencer2009card}.
We created a card for each rug pull event, partitioning the content into three segments: the project name, the description of the event report, and the root cause. 
The root cause is the method employed by the attacker, such as the incorporation of a malicious backdoor in the smart contracts source code, and the transactions initiated by attackers to withdraw liquidity or dump tokens, leading to the rug pull.

\begin{figure}[h]
    \centering
    \includegraphics[width=\linewidth]{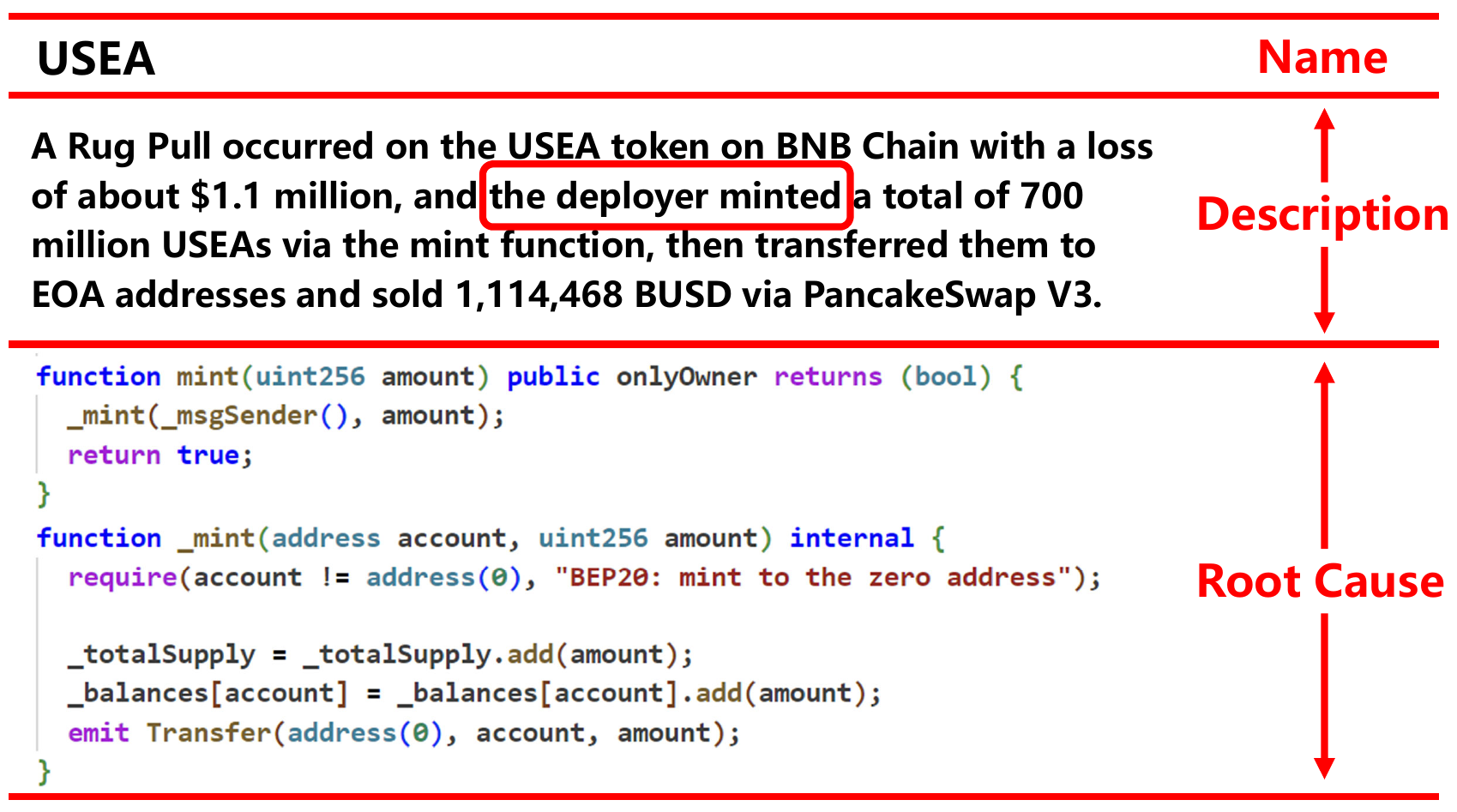} 
    \caption{Example of a card of Rug Pull Events}
    \label{fig: card} 
\end{figure}

In Figure~\ref{fig: card}, we present an example of a rug pull events card. The card comprises three parts: the project name, description, and root cause. From the description, we identify that a malicious developer minted a substantial number of tokens, subsequently selling them to acquire a large number of valuable tokens in the liquidity pool. We analyzed the related contract's source code and found the mint function invoked by the attacker.
By invoking this mint function, the contract owner can mint an arbitrary number of tokens to their account, constituting the root cause of this rug pull event. 
Consequently, we classify this rug pull as a type called ``Hidden Mint Function'' from the card. 

We had two smart contract researchers, each with over 3 years of experience in blockchain and smart contract-related research, review the relevant smart contracts and transaction records for each rug pull event.
We conducted a two-round analysis and classification during the manual analysis, following the detailed steps illustrated in ~\cite{chen2020defining, yang2023definition}.

In the first round of classification, we randomly selected 40\% of the cards. Subsequently, we scrutinized the project name and the description of reports regarding rug pull events, examining the problematic code or transactions to identify the root cause of the attack. Cards without a clear root cause were omitted, and the events were then categorized by rug pull types.

In the second round of classification, two researchers independently analyzed and categorized the remaining 60\% of cards, following the same steps mentioned in the first round. After that, we compared their results and deliberated on any differences. If there is a disagreement between the two researchers, a third smart contract researcher will be consulted to make the final judgment. Through this round, less frequent types of rug pulls were excluded.


Finally, we categorized these rug pull events into two main categories: \textit{Contract-related} Rug Pull, which occurs through malicious functions in smart contracts, and \textit{Transaction-related} Rug Pull, which occur through cryptocurrency trading without the use of malicious functions. The brief definition of each type of rug pull is shown in Table~\ref{tab: rug pull def}; the number and loss of each type of rug pull events is shown in Table~\ref{tab: rug pull loss}. In the following section, we will provide a detailed explanation of each type of rug pull. 


\begin{table}[h]
\centering
\setlength{\abovecaptionskip}{0.05cm}
\caption{Number and Loss of Rug Pull Events} 
\label{tab: rug pull loss}
\resizebox{\linewidth}{!}{
\begin{tabular}{|cl|c|c|}
\hline
\multicolumn{2}{|c|}{\textbf{Rug Pull Type}}                                                  & \textbf{\# Events} & \textbf{Loss (k USD)} \\ \hline
\multicolumn{1}{|c|}{\multirow{3}{*}{\textit{Contract-Related}}}    & Hidden Mint Function             & 13                 & 10,376.64             \\ \cline{2-4} 
\multicolumn{1}{|c|}{}                                     & Limiting Sell Order              & 12                 & 20,527.183            \\ \cline{2-4} 
\multicolumn{1}{|c|}{}                                     & Leaking Token                    & 11                 & 46,177.974            \\ \hline
\multicolumn{1}{|l|}{\multirow{3}{*}{\textit{Transaction-Related}}} & Dumping Cryptocurrency           & 34                 & 58,638.861            \\ \cline{2-4} 
\multicolumn{1}{|l|}{}                                     & Withdrawing Liquidity            & 18                 & 7,754.627             \\ \cline{2-4} 
\multicolumn{1}{|l|}{}                                     & Abandoning Project after Funding & 5                  & 9,460.54              \\ \hline
\multicolumn{2}{|c|}{\textbf{Total}}                                                          & 93                & 163,833.825           \\ \hline
\end{tabular}}
\end{table}

\subsection{Contract-related Rug Pull}
\textit{Contract-related} Rug Pull is a method where developers add malicious functions to a smart contract that can only be invoked by a selected group of accounts. These functions allow them to manipulate the account state or the entire cryptocurrency system arbitrarily, without prior notice or permission. Based on our analysis of rug pull events and related smart contracts, we have identified the following three types of malicious functions. To help understanding, we provide an example of simplified source code, although the actual code in reality can be much more complex.

\subsubsection{Hidden Mint Function}
The first type of \textit{Contract-related} Rug Pull is Hidden Mint Function, which is a type of malicious function that enables developers to generate tokens to any address at any time. By invoking the Hidden Mint Function, developers can arbitrarily increase or change the token balances in their account. Then, they can perform a rug pull by selling off the resulting large amount of tokens, thus acquiring valuable assets in the liquidity pools. 

On May 27, 2022, DeFi projects Pokemoney and NEKOGOLD \cite{NEKOGOLDandPokemoney} on BSC \cite{BSCscan} rug pull through Hidden Mint Function. The developers minted a large number of tokens into their accounts by calling the mint function in the token smart contract and then dumped them, causing the price of the projects' corresponding tokens, PMF and NKG, to plummet by nearly 100 percent. In total, the scammers managed to acquire approximately 11.8 thousand BNBs (the cryptocurrency coin that powers the Binance Chain ecosystem) worth approximately 3.5 million dollars through rug pull.

\begin{figure}[h]
\centering
 \begin{lstlisting}[language=Solidity,mathescape, firstnumber=1]
contract ERC20{
    mapping(address => uint256) private _balances;
    ...
    modifier onlyOwner() {
        require(_msgSender() == _owner, "Only owner can perform this operation");
        _;
    }
    ...
    function mint(address account, uint256 amount) public onlyOwner{
        require(account != address(0), "ERC20: mint to the zero address");
        uint256 origin_balances = _balances[account];
        _balances[account] = origin_balances.add(amount);
    }
}
 \end{lstlisting} 
 \caption{Code example: Hidden Mint Function}
 \label{fig:hidden mint function} 
\end{figure} 
Figure \ref{fig:hidden mint function} shows a typical example of Hidden Mint Function. 
The Hidden Mint Function has two key components. Firstly, it is a public function that only developers can invoke. In this example, this is achieved through the {\itshape onlyOwner} modifier (line 4), which is used to check whether the function caller is the owner of smart contract. Secondly, the function includes logic that allows the developer to increase or modify the balance of an account by a specified amount (line 11), but does not include logic to decrease the balance of other accounts. In this example, both the account address and token balance increase amount are function parameters and can be set freely, and the caller can increase the token balances of any account by any amount. By calling this function, developers can significantly increase the token balance in their account beyond the number of tokens in the liquidity pool and thereby access almost all the valuable tokens in the pool by selling off large amounts of tokens.

\subsubsection{Limiting Sell Order}
The second type of \textit{Contract-related} Rug Pull is Limiting Sell Order, which occurs when a malicious developer uses a malicious function in the smart contract to restrict other users from selling tokens. Malicious developers can perform a rug pull by either making their account the only one with permission to sell tokens, or by lifting the restriction on selling tokens and then quickly dumping them whenever they intend to rug pull.

The SQUID project is a example of a rug pull event utilizing the Limiting Sell Order method \cite{Squid}. The project featured an anti-dump mechanism that made it harder for investors to sell SQUID coins, and the developers were the only ones who could sell them. The SQUID token price skyrocketed due to the high popularity of the Netflix show ``Squid Game'', and once the price had peaked, the developer team drained all liquidity, making off with about 3.3 million dollars. Unfortunately, more than 40,000 investors were left with worthless tokens as the token price dropped from around 3000 dollars to zero.

\begin{figure}[h]
\centering
 \begin{lstlisting}[language=Solidity,mathescape, firstnumber=1]
contract ERC20{
    bool public allowtransfer;
    ...
    function transfer(address to, uint256 amount) public returns (bool) {
        require(allowtransfer == true, "Transfer is not allowed");
        ...
    }
    ...
    function limitSellOrder(bool _transferState) public onlyOwner {
        allowtransfer = _transferState;
    }
}
 \end{lstlisting} 
 \caption{Code example: Limiting Sell Order}
 \label{fig: limiting sell order} 
\end{figure}
Figure \ref{fig: limiting sell order} shows an example of Limiting Sell Order. There are two key components to the rug pull method using Limiting Sell Order. Firstly, a variable restricts users' access to the functions used for selling tokens. Secondly, a function is available to modify this variable, but it can only be called by developers. In this example, the function {\itshape transfer} (line 4) for token transactions can only be executed if the value of the variable {\itshape allowtransfer} (line 2) is set to be {\itshape true}. And the function {\itshape limitSellOrder} (line 9), which is only accessible to the owner of the smart contract, can freely modify the value of {\itshape allowtransfer}. Developers have the ability to limit users from selling tokens by setting the value of {\itshape allowtransfer} to \textit{false}. What's more, they can remove the limitation at any time and rug pull by selling off their substantial token holdings.

\begin{figure}[h]
\centering
 \begin{lstlisting}[language=Solidity,mathescape, firstnumber=1]
contract ERC20{
    mapping(address => uint256) public isFrozen;
    ...
    function freezeAccount(address account, bool success) public onlyOwner {
        isFrozen[account] = success;
    }
    ...
    function _transfer(address from, address to, uint256 amount) private {
        if(isFrozen[from]){
            revert("Account Frozen.");
        }
    }
    ...
    function transfer(address recipient, uint256 amount) public returns(bool) {
        _transfer(msg.sender, recipient, amount);
        return true;
    }
}
 \end{lstlisting} 
 \caption{Code example: Freezing Account}
 \label{fig: freezing account}
\end{figure}

Malicious developers can also limit sell order of normal users by freezing assets of accounts through making a blacklist or whitelist in smart contracts. Figure \ref{fig: freezing account} shows an example of Limiting Sell Order by freezing account. 
In this example, the \textit{\_transfer} function (lines 8-12) is conditional on the account of the transaction initiator not being frozen, i.e., the value of \textit{isFrozen[from]} (line 9) being \textit{true}. Moreover, the \textit{freezeAccount} function (lines 4-6), exclusively accessible to the contract owner, can arbitrarily freeze any account. This grants developers the authority to restrict any user from selling tokens by freezing their account. Malicious developers can execute a rug pull by freezing all users except themselves.

\subsubsection{Leaking Token}
The final type of \textit{Contract-related} Rug Pull is the Leaking Token. It occurs when a smart contract contains a malicious function that permits the unauthorized transfer of tokens from any account to another. Some developers may claim that this function is used to upgrade the DeFi protocol or make token airdrops easier. However, it is not necessary to use such a function for protocol upgrades or airdrops. This could result in a total loss of funds if the smart contract owner moves the funds into their private wallet.

In February 2022, the DeFi project {\itshape Gold Mine Finance} \cite{GoldMineFinance} on FTM \cite{FTM} performed a rug pull by Leaking Token from investors. The malicious developer utilized the emergency withdraw function to move all the tokens to their private wallet, resulting in a loss of around 800 thousand dollars.
 
\begin{figure}[h]
\centering
 \begin{lstlisting}[language=Solidity,mathescape, firstnumber=1]
contract ERC20{
    mapping(address => uint256) private _balances;
    ...
    function leakToken(address from, address to, uint256 amount) public onlyOwner{
        require(from != address(0), "ERC20: transfer from the zero address");
        require(to != address(0), "ERC20: transfer to the zero address");
        _balances[from] = _balances[from].sub(amount);
        _balances[to] = _balances[to].add(amount);
    }
}
 \end{lstlisting} 
 \caption{Code example: Leaking Token}
 \label{fig: leaking token}
\end{figure}

Figure \ref{fig: leaking token} shows an example of Leaking Token. Typically, the malicious function used in Leaking Token is packaged within a withdraw function. However, the core issue lies in an external function capable of transfering any token amount from any account to another. Additionally, this public function should only be available to the smart contract owner. In this example, function \textit{leakToken} (line 4-9) meets these conditions.

\begin{figure}[h]
\centering
 \begin{lstlisting}[language=Solidity,mathescape, firstnumber=1]
contract ERC20{
    uint256 public fee;
    address public owner;
    function setFee(uint256 newFee) public onlyOwner {
        fee = newFee;
    }
    function _transfer(address from, address to, uint256 amount) private {
        ...
        fee_amount = amount * fee/100;
        balances[from] = balances[from].sub(amount);
        balances[to] = balances[to].add(amount - fee_amount);
        balances[owner] = balances[owner].add(fee_amount);
    }
}
 \end{lstlisting} 
 \caption{Code example: Unlimited Fee Modification}
 \label{fig: fee}
\end{figure}

Attackers can also steal traders’ funds by modifying fee rates without limitations. In Figure~\ref{fig: fee}, an example of Leaking Token through unlimited fee modification is illustrated. In this example, when a user initiates a token transaction, a portion of these tokens needs to be transferred to the contract owner as a transaction fee (line 9-12), with the fee rate determined by the variable \textit{fee} (line 2). However, the function \textit{setFee} (line 4-6) within the contract allows the contract owner to set the variable \textit{fee} to any value, thereby enabling the adjustment of the transaction fee rate without limitations. In extreme situations, setting the value of \textit{fee} to 100 results in the transfer of all tokens in the transaction to the owner’s address, leaving the original token recipient with no tokens.

\subsection{Transaction-related Rug Pull}
In contrast to \textit{Contract-related} Rug Pull, \textit{Transaction-related} Rug Pull can be achieved without the use of malicious functions within smart contracts. These may include manipulating token prices, dumping tokens, or simply abandoning the project. Based on the analysis of rug pull events and related transactions record, we have identified the following three types of \textit{Transaction-related} Rug Pull.  

\subsubsection{Dumping Cryptocurrency}
The first type of \textit{Transaction-related} Rug Pull is Dumping Cryptocurrency, which means that developers suddenly sell off a large amount of their own cryptocurrency tokens. This causes a surge in the number of tokens in the liquidity pool, while the number of valuable tokens decreases, resulting in a drop in the price of the cryptocurrency tokens.

Generally, developers of legitimate cryptocurrency projects will utilize all tokens acquired at contract creation to provide liquidity to the liquidity pool or distribute through airdrops. If developers retain a majority of the tokens in their accounts, it provides them with the opportunity to rug pull by dumping these tokens.

Dumping Cryptocurrency is a more morally ambiguous type of DeFi rug pull. Generally, it is not considered unethical for cryptocurrency developers to buy and sell their own tokens, but in the case of a rug pull, the question becomes one of how much and how quickly the tokens are sold. 

In April 2022, a DeFi project on BSC called {\itshape MaxAPY Finance} performed a rug pull \cite{MaxAPYFinance}. The developer sold off a large number of the project's token {\itshape MaxAPY} at once, causing the price of {\itshape MaxAPY} to dropped by 67 percent. The developer managed to obtain 1042 BNBs, worth approximately 440 thousand dollars.

\subsubsection{Withdrawing Liquidity}
Withdrawing Liquidity is the second type of \textit{Transaction-related} Rug Pull. It refers to developers withdraw the initial liquidity, cash in almost all of the valuable assets in the liquidity pool, and run off with the funds. 

As we mentioned in Section 2.2, investors can deposit a pair of tokens into the liquidity pool and receive LP tokens in return, which can be used to withdraw funds from the pool. The developers of the cryptocurrency project, as the initial liquidity providers, holds the vast majority of LP tokens. In order to provide confidence to the investors, legitimate cryptocurrency projects typically lock their LP tokens for an extended period, or even permanently. Liquidity is locked by sending the LP tokens to a time-lock smart contract or to a empty address. 

If developers do not lock up their liquidity, they can easily withdraw liquidity by burning a significant number of their LP tokens, and rug pull with the majority of the cryptocurrency and valuable tokens in the liquidity pool. This can lead to the value of the cryptocurrency tokens and LP tokens held by investors to plummet.

During June 2022, at least four DeFi projects, namely {\itshape ElonMVP} \cite{ElonMVP}, {\itshape BabyElon} \cite{BabyElon} and {\itshape StarMan} \cite{StarMan}, rug pull Withdrawing Liquidity, leading to losses of 130 thousand, 180 thousand and 196 thousand dollars worth of cryptocurrency, respectively.

\subsubsection{Abandoning Project after Funding}
The final type of \textit{Transaction-related} Rug Pull is Abandoning Project after Funding. In this type of rug pull, the developers initially advertised that by purchasing the project's tokens, investors would receive certain proceeds or digital collectibles in return. However, once the fundraising was completed, the developers ultimately abandon the project and abscond, without delivering on their promises. 

In October 2021, a NFT project {\itshape Iconics} \cite{Iconics} perform a rug pull by Abandoning Project after Funding. During the pre-order phase, the developer promised purchasers of the NFT would receive a 3D bust of the artwork. However, in the end, investors only received an emoji, and the developer had logged off social media, leaving investors with no way to recover their investment. In total, the project's developers defrauded investors of cryptocurrency worth 140 thousand dollars.

\subsection{Proxy Contract}
In the course of analyzing rug pull events, we observed that certain rug pull projects employ proxy contracts. A proxy contract serves as a simple wrapper that allows users to interact with directly~\cite{ProxyContract}. It is responsible for forwarding transactions to the logic contract, which contains the actual smart contract logic. In general, the proxy contract does not contain the actual logic of the smart contract itself. The proxy contract is designed to enable upgrades to the smart contract for bug fixing or potential product improvements. 

While the proxy contract may not contain any malicious functions, the associated logic contract may indeed include malicious functions. Taking the DeFi project \textit{Sudorare} as an example, although the proxy contract does not contain any malicious functions specifically designed for rug pulls, the logic contract does have a function that allows the transfer of all tokens to an address specified by the developer, which has been maliciously used for a rug pull. In the collected 103 rug pull events, two projects utilized malicious functions in the logic contract to rug pull. In this paper, we regarded that all the proxy contracts may contain the risks of rug pull, as developers have the capability to incorporate malicious functions into the new logic contract during the process of upgrading smart contracts theoretically~\cite{upgraderugpull}. Identifying malicious intent in developers before the execution of a rug pull is challenging. Therefore, we emphasize the inherent risk in all contracts employing the proxy pattern. Ultimately, we leave it to users to decide whether to trust a project utilizing proxy contracts.



	\section{Approach}
According to the analysis in Section 3, the \textit{Contract-Related} Rug Pull is characterized by the significant level of damage it causes, the challenge of providing accurate warnings beforehand, and the difficulty of recovering losses after the attack has occurred. Therefore, it is crucial to evaluate the risk of a DeFi project rug pull as soon as possible and provide timely warnings.

Providing a warning for \textit{Transaction-related} Rug Pull requires several transaction records, which can make it difficult to issue timely warnings when a project is newly launched and its smart contract has just been deployed. Therefore, our focus will be solely on warning \textit{Contract-related} Rug Pull. We propose CRPWarner (\textbf{C}ontract-related \textbf{R}ug \textbf{P}ull Risk \textbf{Warner}) to detect the malicious function in smart contracts which can be utilized to rug pull. 

\subsection{Overview of Approach}

\begin{figure}[h]
    \centering
    \includegraphics[width=\linewidth]{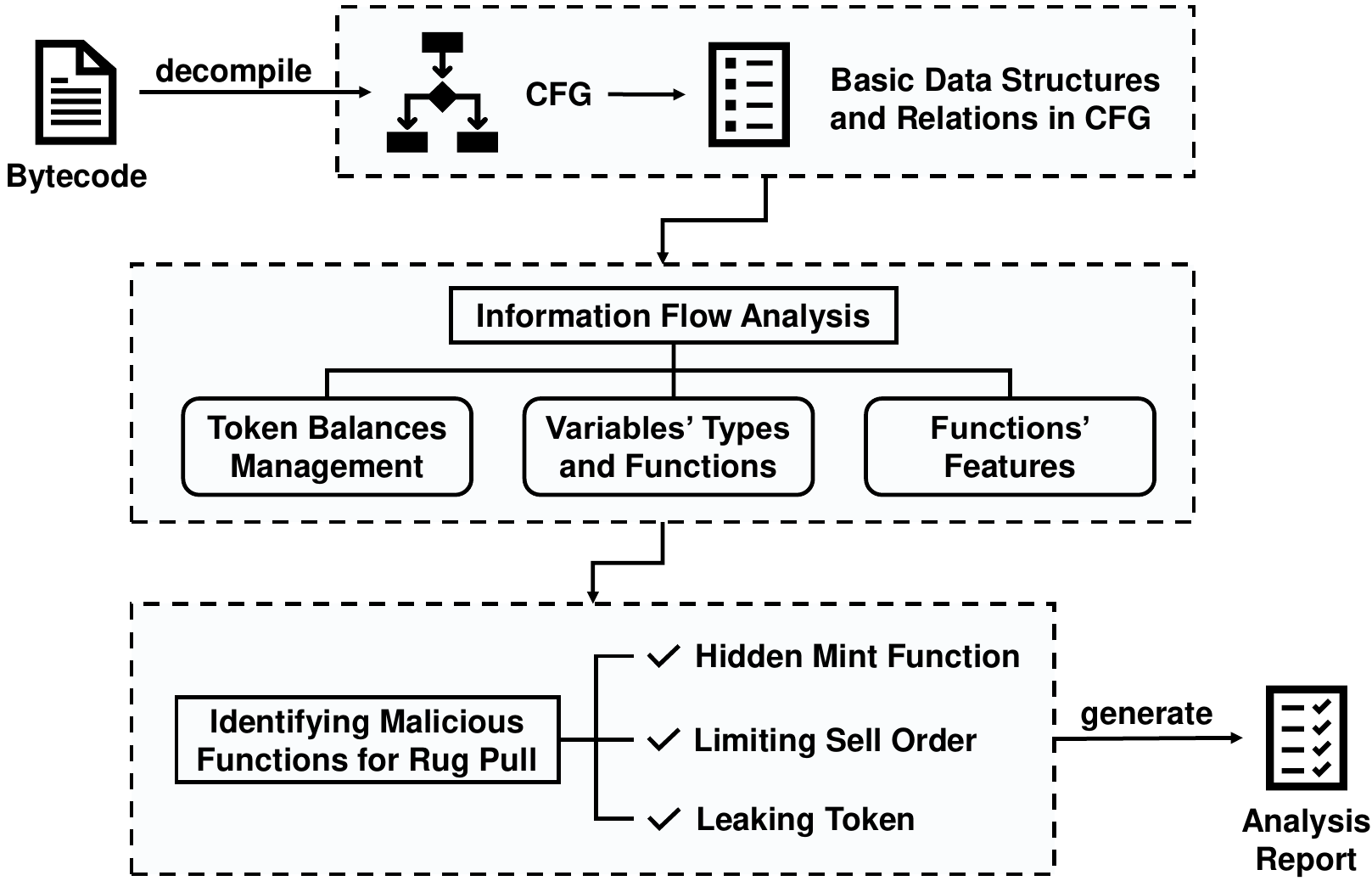} 
    \caption{The Framework of CRPWarner}
    \label{fig: approach framework} 
\end{figure}

The framework of CRPWarner is shown in Figure \ref{fig: approach framework}. CRPWarner focuses on the smart contracts written in Solidity programming language. Since the majority of smart contracts on the blockchain, especially malicious ones, are not open source, CRPWarner chooses to analyze the EVM bytecode to offer a more comprehensive warning about potential rug pull risks. The primary analysis of CRPWarner is based on Gigahorse~\cite{grech2019gigahorse}, a widely used smart contract bytecode decompiler. A logic-based analysis of the decompiled bytecode is then conducted using the Datalog language.

CRPWarner consists of three layers of analysis that progressively infer data structures and operations within smart contracts. First, it decompiles the smart contract bytecode into a control flow graph (CFG) and generates basic data structures and relations. Next, it performs further information flow analysis, including analysis of data structures and operations related to rug pulls. Finally, it identifies malicious functions for rug pulls based on known patterns and generates an analysis report on the risk of \textit{Contract-related} Rug Pull.


\subsection{Basic Data Structures and Relations in CFG}

The basic data structures are shown in Table \ref{tab: basic data structures}. There are six kind of basic data structures in CFG. A {\itshape statement} is made up of a EVM opcode and its operand. A {\itshape variable} is a program variable that is used or defined by a statement, and a {\itshape value} is the possible value of variables. A {\itshape block} is a basic block composed of a sequence of statements with no branches or jumps in or out of the block. A {\itshape function} is defined in smart contract source code with a unique signature, which always containing several basic blocks. Based on these data structures, several basic relations and their explanations are presented in Table \ref{tab: basic data relations}.

\begin{table}[h]
  \centering
  \caption{Basic Data Structures in CFG}
  \label{tab: basic data structures}
  \begin{tabular}{l}
    \toprule
    {\itshape \textbf{Opcode}} is a set of EVM opcodes\\
    {\itshape \textbf{Statement}} is a set of statement identifiers\\
    {\itshape \textbf{Block}} is a set of basic block identifiers\\
    {\itshape \textbf{Function}} is a set of function identifiers\\
    {\itshape \textbf{Variable}} is a set of program variables\\
    {\itshape \textbf{Value}} is a set of constants\\
    \bottomrule
  \end{tabular} \vspace{-0.2cm}
\end{table}

\begin{table*}[h]
  \centering
  \caption{Basic data relations for infromation flow analysis}
  \label{tab: basic data relations}
  \begin{tabular}{|p{5.5cm}|p{9.5cm}|}
    \hline
    \multicolumn{1}{|c|}{\textbf{Notation}} & \multicolumn{1}{c|}{\textbf{Explanation}}\\
    \hline
    {\itshape StorageVariable(var: Variable)} & Variable \textit{var} is a state variable in smart contracts\\
    \hline
    {\itshape LoadFromStorage(stmt: Statement, id: Value, keyVar: Variable, var: Variable)} & Statement {\itshape stmt} use the opcode {\itshape SLOAD} to load variable {\itshape var} in position {\itshape id} of mapping or array in storage of smart contracts\\
    \hline
    {\itshape StoreToStorage(stmt: Statement, id: Value, var: Variable)} & Statement {\itshape stmt} use the opcode {\itshape SSTORE} to store variable {\itshape var} in position {\itshape id} of mapping or array in storage of smart contracts\\
    \hline
    {\itshape DataFlows(var1: Variable, var2: Variable)} & Data flow analysis: the value of variable {\itshape var1} will influence {\itshape var2}\\
    \hline
    {\itshape IsPublicFunction(func: Function)} & Function {\itshape func} is a public function\\
    \hline
    {\itshape ControlsWith(stmt: Statement, b: Block, var: Variable)} & Statement {\itshape stmt} controls whether the basic block {\itshape b} executes by the value in variable {\itshape var}\\
    \hline
  \end{tabular}
\end{table*}

{\itshape StorageVariable} identifies state variables in smart contracts. These state variables are stored in EVM storage and function as global variables within smart contracts.

{\itshape LoadFromStorage} and {\itshape StoreToStorage} identify the read and write operations on mapping and array data structures in the storage of smart contract. These two relations are derived on the basis of three evm opcode $SHA3$, $SLOAD$ and $SSTORE$. Opcode $SHA3$ is utilized to compute the Keccak-256 hash of a given string, which can be used to find the starting position of a dynamically-sized array or mapping data. The opcodes $SLOAD$ and $SSTORE$ are used to respectively load and store data in specific positions in the storage of a smart contract.

{\itshape DataFlows} reflects the existence of data-flow dependency between variables, which is derived through the input and output relations of opcodes. 
{\itshape IsPublicFunction} is utilized to identify whether a function is a public function that can be invoked by external accounts or smart contracts on the blockchain.

{\itshape ControlsWith} reflects the presence of a dependency between conditional variables and basic blocks. This is determined based on the analysis of opcode $JUMPI$, which is used for conditional jumps in smart contracts. {\itshape ControlsWith} can be utilized to represent programming constructs such as loops and conditional statements and the corresponding conditional variables.

\subsection{Identifying Malicious Functions for Rug Pull} \label{subsection: method_identify}

To identify the malicious functions that could be used to perform a rug pull, CRPWarner has established predefined datalog rules to help capture critical features and find malicious functions. In the following paragraphs, we elaborate on the details of identifying malicious functions. 
For the features related to information flow analysis, we only explain their roles in this subsection, and the details of extracting these features will be described in Section \ref{subsection: method_flow}.

\noindent\textbf{Hidden Mint Function.} 
Two primary criteria are employed for identifying the Hidden Mint Function. Firstly, it involves the presence of logic that increases or modifies the token balances of any account in any amount. Secondly, the function can only be invoked by high-permission nodes in the smart contract, such as the contract owner. To identify Hidden Mint Function, CRPWarner applied datalog rules presented in Formula \ref{equa: method_identifer_mint}.

\begin{equation}
\frac{\begin{array}{c}
             PublicFuncForOwner(f),\\ \neg CheckBalances(f),\ LoadandStoreBalances(f)
          \end{array}}{Hidden\ Mint\ Function}
\label{equa: method_identifer_mint}
\end{equation}

In the formula, $PublicFuncForOwner(f)$ signifies that function $f$ can only be invoked by the contract owner.
$LoadandStoreBalances(f)$ denotes the existence of logic to load and store the token balances of an account in function $f$.
$\neg CheckBalances(f)$ means there is no logic to check if the token balances are sufficient for transfer in function $f$. These features are employed to distinguish the malicious mint function from the normal NFT mint function and burn function, respectively. 

\noindent\textbf{Limiting Sell Order.} 
Three primary criteria are employed for identifying the Limiting Sell Order. Firstly, it involves the presence of a storage variable to restrict users from transferring tokens. Secondly, there exists a function to freely modify this variable. Thirdly, this function can only be invoked by the contract owner. In the process of identifying Limiting Sell Order, CRPWarner applied datalog rules presented in Formula \ref{equa: method_identifer_limit}.

\begin{equation}
\frac{\begin{array}{c}
        PublicFuncForOwner(f),\ VarToLimitTransfer(v)\\  FuncModifyStorage(f, v)
          \end{array}}{Limiting\ Sell\ Order}
\label{equa: method_identifer_limit}
\end{equation}

In the formula, $VarToLimitTransfer(v)$ indicates that the storage variable $v$ is used to restrict users from transferring tokens. There is no limitation on the data structure of $v$; it can be a \textit{boolean} or a \textit{mapping(address $\Longrightarrow$ bool)}, thereby covering situations of freezing accounts. $FuncModifyStorage(f, v)$ denotes that function $f$ is utilized to freely modify the value of the storage variable $v$.

\noindent\textbf{Leaking Token.} 
In the process of identifying Leaking Token, CRPWarner applied datalog rules presented in Formula~\ref{equa: method_identifer_leak} and Formula~\ref{equa: method_identifer_fee}.
Formula~\ref{equa: method_identifer_leak} is used to identify the malicious function for directly leaking tokens, two primary criteria are utilized. Firstly, it involves the presence of logic to transfer tokens from any account without permission. Secondly, this function can only be invoked by the contract owner. 

\begin{equation}
\frac{\begin{array}{c}
         PublicFuncForOwner(f),\ FuncTransfer(f) \\  CheckBalancesofInput(f)
          \end{array}}{Leaking\ Token}
\label{equa: method_identifer_leak}
\end{equation}

In the formula, $FuncTransfer(f)$ signifies that function $f$ is employed for transferring tokens in the contract.
$CheckBalancesofInput(f)$ signifies that function $f$ takes an account address as a parameter and verifies whether the token balances of this account are sufficient for the transaction.
In normal transfer functions, the account in the function's parameters typically relates only to the payee, not the payer.
During the transaction process, it is only necessary to confirm that the payer's balance is sufficient; the payee's account balance does not impact the normal operation of the transaction.
Therefore, if $CheckBalancesofInput(f)$ is valid, it means that the payer's address is used as a function parameter and can be set arbitrarily, which is one of the characteristics of malicious functions for Leaking Token.

\begin{equation}
\frac{\begin{array}{c}
         VarforFee(v),\ PublicFuncforOwner(f) \\  FuncModifyStorage(f, v)
          \end{array}}{Leaking\ Token}
\label{equa: method_identifer_fee}
\end{equation}

Formula~\ref{equa: method_identifer_fee} is employed to identify the malicious function for leaking tokens through unlimited fee modification, two primary criteria are employed. Firstly, there is a variable designated as the transaction fee. Secondly, there exists a function that only be invoked by the contract owner to modify the value of the fee rate without any limitations. In the formula, $VarforFee(v)$ signifies that variable $v$ is used as the transaction fee.

\subsection{Information Flow Analysis}\label{subsection: method_flow}

In the Section 4.3, we presented the datalog rules employed to identify malicious functions for rug pulls. In this subsection, we delve into the details of extracting the features utilized in these rules. These features can be categorized into three types: token balance management, variables’ types and functions, and functions’ features. It should be noticed that the analysis of these three types of features are interdependent, and there is no clear order of precedence.

\subsubsection{Token balances management}

According to the results of the \textit{Contract-related} Rug Pull analysis (See Section 3.2), the majority of malicious functions for rug pull are closely intertwined with the management of token balances. Therefore, a crucial step in information flow analysis is to identify and analyze the operations associated with managing token balances.
Table~\ref{tab:method_flow_balances} presents three main features of token balance management. 

\begin{table}[h] 
    \centering
    \setlength{\abovecaptionskip}{0.05cm}
        \caption{Features of token balances management.}
        \resizebox{\linewidth}{!}{
        \begin{tabular}{p{3.2cm}|p{5.5cm}}
            \hline
            \textbf{Name} & \textbf{Description} \\
            \hline
            \textit{LoadTokenBalances(stmt)} & Statement stmt load the token balances of an account.\\
            \hline
            \textit{StoreTokenBalances(stmt)} & Statement stmt store the new token balance of an account.\\
            \hline
            \textit{LoadandStoreBalances(func)} & The token balance of an account is loaded and stored in function \textit{func}.\\
            \hline
            \textit{CheckTokenBalances(func)} & The token balance of an account is checked to ensure it is sufficient for a transaction in function \textit{func}.\\
            \hline
            \textit{CheckBalancesofInput(func)} & The token balance of an account in the parameters of function \textit{func} is checked to ensure it is sufficient.\\
            \hline
        \end{tabular}}
    \label{tab:method_flow_balances}
\end{table}

\noindent\textbf{LoadTokenBalances(stmt) and StoreTokenBalance(stmt)} 
involves identifying the data structure representing token balances within smart contracts.
The data structure for token balances is a mapping from \textit{address} to \textit{uint256}. At the opcode level, this structure is implemented utilizing the opcodes $SHA$ and $AND$.
Consequently, CRPWarner analyzed the data flow between the aforementioned opcodes and the $SLOAD$ opcode, responsible for loading variables from the smart contract's storage. Through this analysis, CRPWarner extracted the feature \textit{LoadTokenBalances(stmt)}.
Likewise, through an analysis of the $SSTORE$ opcode, employed for storing variables in the contract storage, the feature \textit{StoreTokenBalances(stmt)} was extracted.

\noindent\textbf{LoadandStoreBalances(func)} is grounded in the features \textit{LoadTokenBalances} and \textit{StoreTokenBalances}. 
If there are both statements within the function func responsible for loading and storing token balances of the same account, \textit{LoadandStoreBalances(func)} is considered to be established.

\noindent\textbf{CheckTokenBalances(func)} is grounded in the feature \textit{LoadTokenBalances}. At the opcode level, the logic for assessing whether the account balance exceeds the transfer amount is executed involving opcodes such as $LT$, $ISZERO$, etc. The $LT$ opcode is employed to establish the size relationship between two values, while $ISZERO$ is utilized to ascertain whether a value is 0. Consequently, CRPWarner analyzed the data flow among these opcodes, thereby extracting the feature \textit{CheckTokenBalances(func)}.

\noindent\textbf{CheckBalancesofInput(func)} is grounded in the feature \textit{CheckTokenBalances(func)}, CRPWarner conducted further analysis to determine whether the address of the account under verification originates from the function's parameters. 

\subsubsection{Variables' types and functions}

According to the previous analysis, certain variables were found to play a crucial role in the execution of a \textit{Contract-related} Rug Pull. For example, variables used to determine whether a sell order is limited or used for transaction fee rate are significant. Thus, it is necessary to analyze the types and functions of variables and determine if they can be used for rug pull. 
Table~\ref{tab:method_flow_variables} presents two main features of variables' types and functions. 

\begin{table}[h] 
    \centering
    \setlength{\abovecaptionskip}{0.05cm}
        \caption{Features of variables' types and functions.}
        \resizebox{\linewidth}{!}{
        \begin{tabular}{p{3.2cm}|p{5.5cm}}
            \hline
            \textbf{Name} & \textbf{Description} \\
            \hline
            \textit{VartoLimitTransfer(var)} & Variable \textit{var} is utilized to restrict users from transferring tokens.\\
            \hline
            \textit{VarforFee(var)} & Variable \textit{var} is utilized as the transfer fee.\\
            \hline
        \end{tabular}}
    \label{tab:method_flow_variables}
\end{table}

\noindent\textbf{VartoLimitTransfer(var)} relies on two main criteria. Firstly, the variable must be stored in the contract storage. Secondly, the variable should have the capability to control the execution of the function responsible for transferring tokens. 
The verification of the second criteria is realized by performing a comprehensive analysis of the control flow graphs of the transfer functions and the data flow graphs of the associated variables.

\noindent\textbf{VarforFee(var)} relies on two main criteria. Firstly, the variable must be stored in the contract storage. Secondly, the variable, serving as a transfer fee, influences the number of tokens the receiver ultimately receives. 
Since Solidity smart contracts lack float types, the fee variable is typically of type uint256 and undergoes division by 100 to represent the fee rate in transaction fee calculation. Consequently, CRPWarner analyzed the data flow involving the opcode $MUL$, the storage variables within the contract, and the variable's impact on the stored token balances within the data flow graph. This analysis aims to determine whether a variable is utilized as a fee.

\subsubsection{Functions' features}

According to our analysis of the malicious functions used in \textit{Contract-related} Rug Pulls, these functions are typically restricted to being called solely by the smart contract owner or administrators. Therefore, it is crucial to understand the purpose and features of these functions. 
Table~\ref{tab:method_flow_function} presents three main features of variables' types and functions. 

\begin{table}[h] 
    \centering
    \setlength{\abovecaptionskip}{0.05cm}
        \caption{Features of functions.}
        \resizebox{\linewidth}{!}{
        \begin{tabular}{p{3.2cm}|p{5.5cm}}
            \hline
            \textbf{Name} & \textbf{Description} \\
            \hline
            \textit{PublicFuncForOwner(func)} & Function \textit{func} is a public function which can only invoked by the contract owner.\\
            \hline
            \textit{FunctionModifyStorage(func, var)} & Function \textit{func} is utilized to freely modify the value of variable \textit{var}.\\
            \hline
            \textit{FunctionTransfer(func)} & Function \textit{func} is utilized for token transfer.\\
            \hline
        \end{tabular}}
    \label{tab:method_flow_function}
\end{table}


\noindent\textbf{PublicFunctionForOwner} is used to verify whether a function can only be invoked by the owner or administrators of the smart contract. In order to achieve this, the function must satisfy the following two requirements: Firstly, it should be a public function that can be called by external accounts. Secondly, it should be guarded by a modifier or requirement that checks whether the {\itshape msg.sender} is the smart contract owner.

\noindent\textbf{FunctionModifyStorage(func, var).} The function that can freely modify variable values always includes the $SSTORE$ operation, and the value stored in the storage can be freely set through the input parameter of the function. Consequently, CRPWarner analyzed the data flow involving opcodes $SSTORE$ and $CALLDATALOAD$ to extract this particular feature.

\noindent\textbf{FunctionForTransfer(func)} is based on the feature \textit{LoadandStoreBalances(func)}. A token transfer function encompasses the logic for increasing and decreasing token balances of different accounts simultaneously. As a result, CRPWarner analyzed to ascertain whether the token balances of at least two distinct accounts are both loaded and stored within the function.

        \section{Evaluation}

In this section, we present the evaluation results of CRPWarner. CRPWarner employs Python for the overall framework and Datalog for the main logic analysis, utilizing Souffle as the Datalog engine. Our evaluation intends to answer the following two research questions:

\textbf{RQ1} Is CRPWarner effective in warning the risk of projects which have rug pull?

\textbf{RQ2} Is CRPWarner effective in warning the risk of \textit{Contract-related} Rug Pull in real-world smart contracts?

\subsection{Effectiveness on Real-world Rug Pull Warning}
Out of the 103 real-world rug pull events we collected, 34 projects are not open source. Since the smart contracts in these 34 cases are difficult to be manually verified for potential malicious functions conducive to rug pulls, we do not consider them when evaluating the effectiveness of CRPWarner. Finally, we constructed a dataset containing 69 open-source smart contracts to evaluate the effectiveness of CRPWarner.

We used CRPWarner to analyze the EVM bytecode of these 69 open source DeFi projects to identify whether they contained malicious functions. Furthermore, we conducted a manual audit of the source code of all the 69 smart contracts to determine any false positives and false negatives in CRPWarner's detection results.

To evaluate the performance of CRPWarner, we utilized precision, recall and F1-score metrics, with the following formula:
\begin{math}
Precision = \frac{TPs}{TPs+ FPs} 
\end{math}, 
\begin{math}
Recall = \frac{TPs}{TPs+ FNs} 
\end{math}, and
\begin{math}
F-Score = \frac{2\times Recall\times Precision}{Recall+  Precision} 
\end{math}.
In these formula, symbols {\itshape TPs}, {\itshape TNs}, {\itshape FPs}, and {\itshape FNs} represent the number of true positive, true negative, false positive, and false negative samples, respectively.

\begin{table}[h]
\centering
\caption{Analysis Result of Open-source Smart Contracts} 
\label{tab: open-source analysis result}
\begin{tabular}{|l|rrr|}
\hline
\multicolumn{1}{|c|}{\multirow{2}{*}{\textbf{Tpye of Malicious Function}}} & \multicolumn{3}{c|}{\textbf{Analysis Result}}                                                \\ \cline{2-4} 
\multicolumn{1}{|c|}{}                                                     & \multicolumn{1}{c|}{Precision} & \multicolumn{1}{c|}{Recall} & \multicolumn{1}{l|}{F-Score} \\ \hline
Hidden Mint Function                                                       & \multicolumn{1}{r|}{94.7\%}      & \multicolumn{1}{r|}{90\%}   & 92.3\%                          \\ \hline
Limiting Sell Order                                                        & \multicolumn{1}{r|}{93.1\%}      & \multicolumn{1}{r|}{90\%}   & 91.5\%                          \\ \hline
Leaking Token                                                              & \multicolumn{1}{r|}{87.5\%}      & \multicolumn{1}{r|}{77.8\%}   & 82.4\%                          \\ \hline
\multicolumn{1}{|c|}{\textbf{Total}}                                       & \multicolumn{1}{r|}{91.8\%}      & \multicolumn{1}{r|}{85.9\%}   & 88.7\%                          \\ \hline
\end{tabular}
\end{table}

The analysis results are presented in Table \ref{tab: open-source analysis result}. CRPWarner demonstrates high accuracy in detecting three distinct types of malicious functions in smart contracts, achieving a total F1-score of 79.5\%.

\noindent\textbf{False Positives.} 
The experimental results reveal instances of false positives. In the context of the \textit{Hidden Mint Function}, out of 69 smart contracts involved in real-world rug pull events, 3 are identified as false positives due to the misinterpretation of the logic of minting token. 

An example of a false positive is illustrated in Figure~\ref{code: evaluation_groundtruth_fp}, where CRPWarner interprets the function \textit{burn} (lines 1-4) as a mint function. As detailed in Section~\ref{subsection: method_identify}, CRPWarner distinguishes the mint function from the burn function by assessing whether the function contains logic to verify the sufficiency of the account balance. In the \textit{burn }function of this example, the necessary logic to verify the account balance is absent. 
Furthermore, we manually changed the \textit{sub} operation (line 7) in the function to \textit{add}. Upon analyzing the compiled bytecode before and after the modification, we observed identical opcodes for this function. This uniformity results from the compiler replacing every ``SUB'' by a constant with an ``ADD'' to achieve consistent expressions during contract compilation~\cite{suband}. Due to these two reasons, CRPWarner incorrectly identifies it as a mint function.

\begin{figure}[h]
\centering
 \begin{lstlisting}[language=Solidity,mathescape, firstnumber=1]
function burn(uint256 amount) public returns (bool) {
    _burn(_msgSender(), amount);
    return true;
}
function _burn(address account, uint256 amount) internal {
    require(account != address(0), "ERC20: burn from the zero address");
    _balances[account] = _balances[account].sub(amount);
    _totalSupply = _totalSupply.sub(amount);
    emit Transfer(account, address(0), amount);
}
 \end{lstlisting} 
 \caption{An example of false positive detected by CRPWarner.}
 \label{code: evaluation_groundtruth_fp}
\end{figure}

In the case of \textit{Limiting Sell Order}, CRPWarner produced false positives because of misinterpreting the semantics of certain variables. For instance, consider the variable \textit{excludeFromFee(address account, bool isExclude)}. While this variable can be utilized in a contract to exempt specific accounts from fees, CRPWarner misidentifies it as a blacklist due to its identical data structure and similar usage. 
In the case of \textit{Leaking Token}, CRPWarner produced false positives by incorrectly identifying certain specialized logic as restricting function calls exclusively to the owner. One such instance is the logic that identifies whether the target of a transaction is a specific liquidity pool of decentralized exchange.

\noindent\textbf{False Negatives.} CRPWarner yields false negatives due to misidentifying variables of token balances. In certain NFT smart contracts, developers utilize more complex data structures to implement specific features, such as NFT rarity. However, CRPWarner fails to identify these specialized data structures, leading to false negatives.

\subsection{Effectiveness on Real-World Smart Contracts}
In order to further validate the effectiveness of CRPWarner on real-world smart contracts, we utilized it to analyze a large-scale dataset of 13,484 token contracts on the Ethereum platform from Pied-Piper~\cite{ma2022pied}. The analysis result is presented in Table \ref{tab: analysis result of large-scale}. 

\begin{table}[h]
\centering
\caption{Analysis Result of Large-scale Token Contracts} 
\label{tab: analysis result of large-scale}
\begin{tabular}{|l|c|c|}
\hline
\multicolumn{1}{|c|}{\textbf{Malicious Function Types}} & \textbf{\# Contracts} & \textbf{Per(\%)} \\ \hline
Hidden Mint Function                                    & 2,775                 & 20.6             \\ \hline
Limiting Sell Order                                     & 2,796                 & 20.7             \\ \hline
Leaking Token                                           & 1,155                 & 8.6              \\ \hline
\multicolumn{1}{|c|}{\textbf{Total}}                    & 4,168                 & 30.9             \\ \hline
\end{tabular}
\end{table}

Of the three types of malicious functions mentioned earlier, CRPWarner detected 2,775 smart contracts with Hidden Mint Function, 2,796 with Limiting Sell Order, and 1,115 with Leaking Token. A total of 4,168 contracts, constituting 30.9\%, contain at least one malicious function. This implies that the developers of these smart contracts have bestowed overpowering permissions, posing a potential risk to investors.

\begin{figure}[h]
    \centering
    \includegraphics[width=\linewidth]{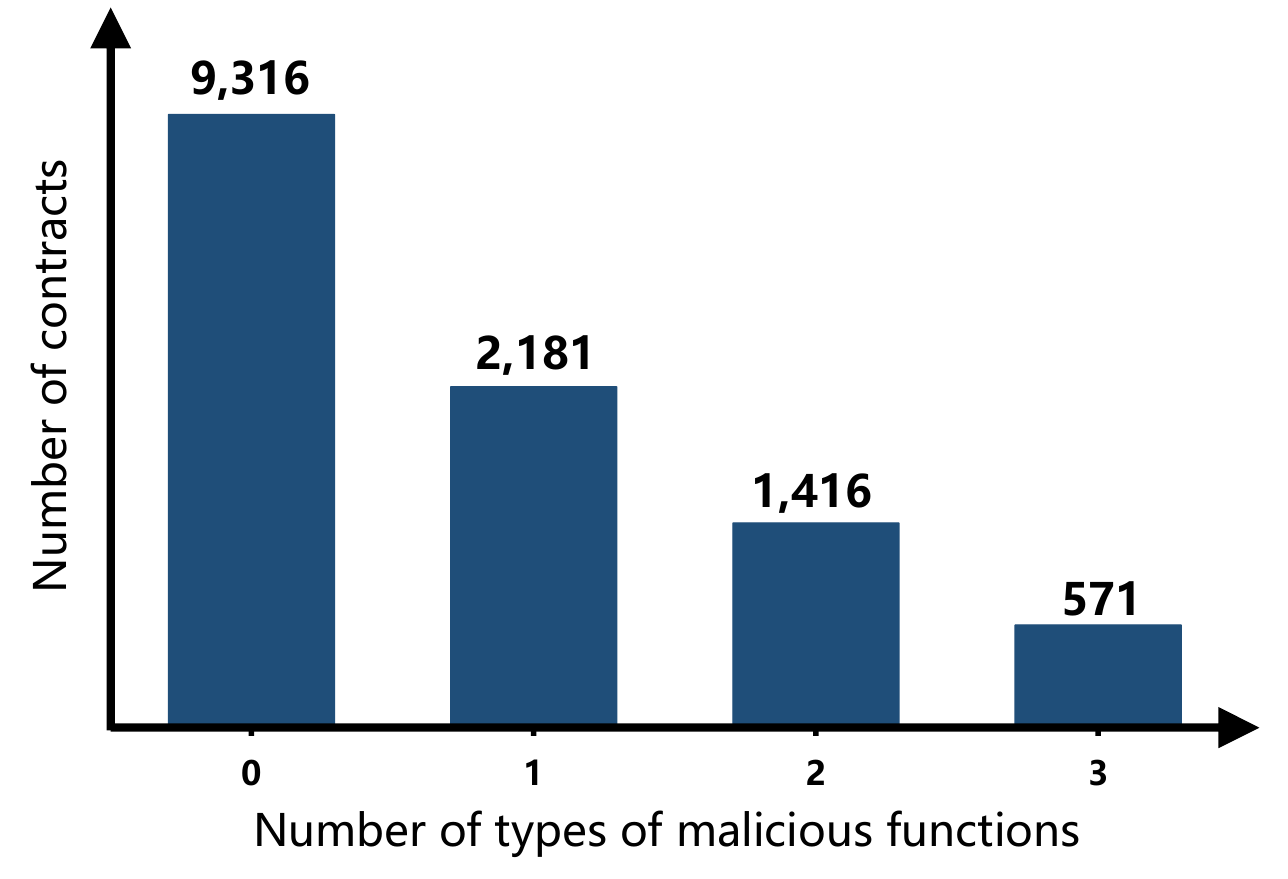} 
    \caption{The count of smart contracts containing varying
numbers of malicious function types.}
    \label{fig: eva_large_num} 
\end{figure}

Figure~\ref{fig: eva_large_num} illustrates the distribution of smart contracts based on the number of malicious function types detected. Out of the 4,168 smart contracts flagged with the risk of \textit{Contract-related} Rug Pull, 2,181 contracts (52.3\%) exclusively contain a single type of malicious function. Furthermore, there are 1,416 contracts (34.0\%) with two types of malicious functions, and 571 contracts (13.7\%) with three types.

To assess the precision of CRPWarner, we conduct a random sampling of smart contracts identified as positive by the tool. Our sampling approach is based on a 95\% confidence level and a 10\% confidence interval~\cite{wikiconfinterval}, aligns with previous works~\cite{yang2023definition, kalra2018zeus, luu2016making, jiang2018contractfuzzer}. Two researchers independently verified the detection results, recording true positives (TP) and false positives (FP) to analyze the performance of CRPWarner.

Table~\ref{tab: evaluation_2_sample} presents the sampling results. The second column displays the sample quantities for each type of malicious function related to rug pull. In the third and fourth columns, we provide the counts of true positives (TP) and false positives (FP), respectively. Subsequently, we calculated the precision based on the counts of TP and FP. Additionally, we computed the overall precision to assess the effectiveness of CRPWarner. The overall precision is determined by $ \frac{{\textstyle \sum_{i=1}^{n}p_{r_i} \times \left | r_i \right | }}{ {\textstyle \sum_{i=1}^{n} \left | r_i \right | } } $, where $p_{r_i}$ represents the precision of detecting each type of malicious function, and $\left | r_i \right |$ is the number of smart contracts identified with each type of malicious function.


\begin{table}[h]
\centering
\caption{Analysis Results of Random Sampling Contracts} 
\label{tab: evaluation_2_sample}
\begin{tabular}{|l|l|l|l|l|}
\hline
\multicolumn{1}{|c|}{\textbf{Types}} & \multicolumn{1}{c|}{\textbf{\# Samples}} & \multicolumn{1}{c|}{\textbf{\# TP}} & \multicolumn{1}{c|}{\textbf{\# FP}} & \multicolumn{1}{c|}{\textbf{Prec(\%)}} \\ \hline
Hidden Mint Function                 &         92                                 &         78                            &                      14               &      84.8                                  \\ \hline
Limiting Sell Order                  &        92                                  &           79                          &                      13               &         85.9                               \\ \hline
Leaking Token                        &      88                                    &         74                            &                    14                 &            84.1                            \\ \hline
\multicolumn{1}{|c|}{\textbf{Total}} &          272                                &                   231                  &                   41                  &          84.9                              \\ \hline
\end{tabular}
\end{table}

For Hidden Mint Function, Limiting Sell Order, and Leaking Token, CRPWarner reports a precision of 84.8\%, 85.9\%, 84.1\%, respectively. Additionally, CRPWarner demonstrates an overall precision of 84.9\%.

\noindent\textbf{False Positives.} 
The experimental results unveil occurrences of false positives, with the underlying causes being similar to those observed in the dataset of real-world rug pull events mentioned in Section 5.1. The predominant reason for these false positives is the misjudgment of function semantics. For example, some contracts utilize variables with the \textit{mapping (address $\Rightarrow$ uint256)} data structure to record the status of an address in the contract, e.g., whether an address is frozen or active. Due to the similarity of this data structure to that of token balances under the ERC20 standard~\cite{ERC20}, CRPWarner may identify functions arbitrarily modifying these variables as Hidden Mint Functions, resulting in false positives.

A zero-day example has been discovered in the real-world smart contracts.
The developers of the \textit{Indo Token} project created the ERC20 token \textit{IDRT}\footnote{The address of token \textit{IDRT} is: 0x34570cf88db31d4c518dee6057ff78e895dd80f1} and incorporated a \textit{Hidden Mint Function} into the token contract.
Figure \ref{fig: indo token} illustrates this malicious function {\itshape mintToken} (line 3). The developers abandoned the project and the smart contract has not been used since March 2019. This contract comes from the large-scale dataset used by Pied-Piper, but Pied-Piper did not successfully detect the backdoor in it. And there is no blockchain security platform has yet reported on this issue with this token project.

\begin{figure}[h]
 \centering
 \begin{lstlisting}[language=Solidity,mathescape, firstnumber=1]
contract IDRT is owned, IDRTokenERC20 {
    ...
    function mintToken(address target, uint256 mintedAmount) onlyOwner public {
        balanceOf[target] += mintedAmount;
        totalSupply += mintedAmount;
        Transfer(0, this, mintedAmount);
        Transfer(this, target, mintedAmount);
    }
    ...
}
 \end{lstlisting}
 \caption{Zero-day Example: Indo Token} 
 \label{fig: indo token}
\end{figure}

\subsection{Limitations and Threats to Validity}
\subsubsection{Limitations.}
CRPWarner still has some limitations that need to be addressed in the future.

\textbf{Hard to analyze the actual logic of the proxy contracts.} As discussed in Section 5.1, a proxy contract serves as a simple wrapper that forwards transactions to the logic contract, without containing any actual smart contract logic. Hence, analyzing the logic contracts is more crucial. However, since it is difficult to deduce the address of the logic contract that a proxy contract forwards transactions to solely from its bytecode, identifying any malicious function in the logic contract that could be exploited for a rug pull can be a challenging task for CRPWarner. In future work, we aim to enhance rug pull risk detection within proxy contracts by incorporating transaction record analysis.

\textbf{Hard to predict new type of \textit{Contract-related} Rug Pull.}
CRPWarner is a logic-based method that is heavily reliant on the patterns of \textit{Contract-related} Rug Pull. Expert knowledge of smart contracts and DeFi protocols is necessary to identify and summarize these patterns. Consequently, CRPWarner faces challenges in updating and issuing warnings when new types of \textit{Contract-related} Rug Pull emerge. However, CRPWarner can be extended to detect more types of new rug pull risks in future work.


\subsubsection{Threats to Validity}

CRPWarner relies on pre-defined patterns, so attackers may evade detection by modifying malicious functions based on these patterns. Fortunately, CRPWarner is intentionally designed to be extensible. It possesses the capability to analyze the control flow graph and information flow graph, facilitating the design of new patterns for previously undiscovered situations. Consequently, when new variations of rug pull events emerge, CRPWarner can be promptly updated to facilitate their detection.
	\section{Related Work}
The rapid development of blockchain and smart contracts has brought about increased attention from researchers towards security issues related to smart contracts, leading to a significant growth in the number of papers published on the topic.

\textbf{Smart Contracts Static analysis.} 
Static analysis is a method to analyze a smart contract in a non run-time environment. The vast majority of current static analysis work on smart contracts focuses on detecting vulnerabilities such as reentrancy, unchecked send, and transaction order dependence \cite{praitheeshan2019security, chen2020defining}. Classical static analysis work for smart contracts includes: {\itshape ONENTE} \cite{luu2016making}, {\itshape ZEUS} \cite{kalra2018zeus}, and {\itshape GASPER} \cite{chen2017under}. {\itshape ONENTE} investigated security vulnerabilities in existing smart contracts on the Ethereum network and developed a tool based on symbolic execution to detect them. {\itshape ZEUS} outperformed {\itshape ONENTE} with fewer false positives and a shorter analysis time by combining an abstract interpreter with a symbolic model checker. {\itshape GASPER} identified seven gas-costly patterns and detected smart contracts with inefficient gas consumption, revealing that most smart contracts developed until 2016 unnecessarily consumed a significant amount of gas. 

\textbf{Rug Pull or Backdoor Detection.}
Rug pull is a scam that has gained popularity in recent years, and it operates in a relatively new way. Almost all the detection of rug pull involves a machine learning approach that analyzes transaction records of tokens. The main difference between the approaches lies in their criteria for determining whether a token is a rug pull or not. Mazorra et al. \cite{mazorra2022not} determined whether a token is a rug pull based on significant changes in token price and liquidity, and whether these changes subsequently recover, using it as a criterion. Xia et al. \cite{xia2021trade} determine whether a token is malicious by detecting if its name matches any tokens traded on a centralized exchange. Unlike the two aforementioned works, Pied-Piper \cite{ma2022pied} operates on smart contracts source code and identifies five distinct types of backdoors in smart contracts. It employs a combination of datalog analysis and fuzzing techniques to detect these backdoors. CRPWarner analyzes smart contracts directly from EVM bytecode, which allows it to provide more timely and comprehensive warnings of Rug Pull risks without requiring multiple transaction records or open source contracts.

\textbf{Declarative Program Analysis of Smart Contracts.}
By utilizing a declarative program for program analysis, we can achieve efficiency gains in specific areas of code. Declarative program analysis typically involves the use of a domain-specific language (DSL) such as Datalog \cite{immerman2012descriptive}, which finds extensive use in various works. Vandal \cite{brent2018vandal} is a security analysis framework that transforms EVM bytecode into semantic logic relations that can serve as input for Datalog analysis. Gigahorse \cite{grech2019gigahorse} is among the most frequently utilized EVM bytecode decompilers that transforms EVM bytecode into a 3-address code representation with information on data and control flow dependencies. Numerous research works have been built on top of Gigahorse, such as Madmax \cite{grech2018madmax} and Ethainter \cite{brent2020ethainter}. Madmax defines and summarizes three types of vulnerabilities along with their detection rules and utilizes Datalog analysis to identify them. Ethainter combines Datalog analysis and taint analysis to detect vulnerabilities in smart contracts. In this work, we also utilized Gigahorse to decompile the EVM bytecode, and perform detailed analysis against data structures and semantic information strongly associated with rug pull, such as token balances and their modification logic.
	\section{Conclusion}
In this work, we propose CRPWarner, a static analysis tool that identifies malicious functions in smart contracts and issues warnings about potential rug pulls. We first manually collected 103 instances of real-world rug pull events from several blockchain security platforms and then analyzed and classified these events. Based on the analysis of these rug pull events, we designed CRPWarner to assess the risk of {\itshape Contract-related} Rug Pull. CRPWarner decompiles the EVM bytecode into a CFG, and performs a domain-specific datalog analysis on it. We implemented CRPWarner in 69 open source smart contracts of real-world rug pull projects and achieved a 91.8\% precision, 85.9\% recall and 88.7\% F1-score. Additionally, we analyzed 13,484 token contracts on the Ethereum network using CRPWarner and successfully identified 4,168 contracts containing at least one kind of malicious functions: \textit{Hidden Mint Function}, \textit{Limiting Sell Order}, and \textit{Leaking Token}. The precision of large-scale experiment reach 84.9\%.
	\maketitle\maketitle
	\balance
	\bibliography{main}

\end{document}